\documentclass[10pt,letterpaper]{article}
\usepackage{opex3}

\begin{document}
\title{Axicon Lens for Coherent Matter Waves}

\author{S. R. Muniz, S. D. Jenkins, T. A. B. Kennedy, D. S. Naik, and C. Raman}

\address{School of Physics, Georgia Institute of Technology, Atlanta, Georgia USA 30332-0430}

\email{srmuniz@gatech.edu}

\homepage{http://www.physics.gatech.edu/chandra} 


\begin{abstract} We have realized a conical matter wave lens. The repulsive potential of a focused laser beam was used to launch a Bose-Einstein condensate into a radially expanding wavepacket whose perfect ring shape was ensured by energy conservation. In spite of significant interactions between atoms, the spatial and velocity widths of the ring along its radial dimension remained extremely narrow, as also confirmed by numerical simulations.  Our results open the possibility for cylindrical atom optics without the perturbing effect of mean-field interactions.
\end{abstract}


\bibliographystyle{osajnl}


\section{Introduction}

Gaseous Bose-Einstein condensates (BECs) present unique possibilities as bright, coherent matter wave sources for applications to coherent atom optics, holography and interferometry \cite{adam94,mori96,atomopt1996,berm97int,meys01book}.  There is much interest in creating novel atom optical elements for focusing and deposition \cite{timp92,mccl94}, coupling of atoms with high efficiency into atom chip guides \cite{henk05special}, collimation of atomic trajectories within an atomic clock to suppress systematic effects \cite{juri05}, and for the exploration of fundamental physics with matter waves \cite{arnd99,koko01}.  However, unlike photonic optics, propagation of matter waves is profoundly affected by the interactions between atoms.  For example, in a gaseous Bose-Einstein condensate, the ``mean-field'' driven expansion destroys the Heisenberg-limited momentum distribution possessed by a condensate while it is trapped, creating a matter wave with a broad spread of velocities that is undesirable for atom-optical applications.  Therefore, it is of great interest to discover methods of controlling and suppressing interaction effects during propagation of a matter wave field and within atom interferometers.  Such approaches include manipulation of the scattering length using Feshbach resonances \cite{inou98}, Raman outcoupling of a BEC \cite{hagl99,robi06}, the use of extremely dilute condensates \cite{lean03} and interferometry with fermionic gases where $s$-wave scattering is forbidden \cite{roat04}.

In this work, we report on the realization of a conical lens, or ``axicon'' for Bose-Einstein condensates (BECs) using the dipole force exerted by a focused, blue-detuned laser beam.  To our knowledge this is the first observation of the far-field pattern of an axicon using neutral atoms.  In our experiment, the atoms expand radially outward in the form of an extremely thin, circular ring in spite of significant interactions between atoms.  This is in marked contrast to the conventional BEC expansion, which is characterized by strong spatial dispersion and a broad final velocity distribution.  We have performed numerical solutions of the Gross-Pitaevskii equation confirming that this interplay between the external and internal forces can be exploited to tailor the expansion of a BEC using light fields.

\section{Experimental Description}

We produce Bose-Einstein condensates of roughly $3 \times 10^6$ sodium atoms in a purely linear quadrupole potential formed by a pair of anti-Helmholtz coils following the procedure in our earlier work \cite{naik05meta}.  To create the BEC, we first use a blue-detuned ``optical plug'' consisting of a far-off resonance laser at 532 nm which has been focused to a $1/e^2$ waist of 40 $\mu$m to prevent Majorana transitions near the magnetic field zero \cite{majo32,naik05,davi95bec}.  The laser beam intensity is ramped to zero within 200 ms after the condensate is produced, resulting in a metastable condensate in a ``linear'' potential formed only by the quadrupole coils, with an axial field gradient of $B_z' = 21$ G/cm.  The predicted radial and axial Thomas-Fermi radii of the condensate are $R_\rho = $18 $\mu$m and $R_z = $10 $\mu$m, respectively.

\begin{figure}[htb]
\centering\includegraphics[width= 0.9\textwidth]{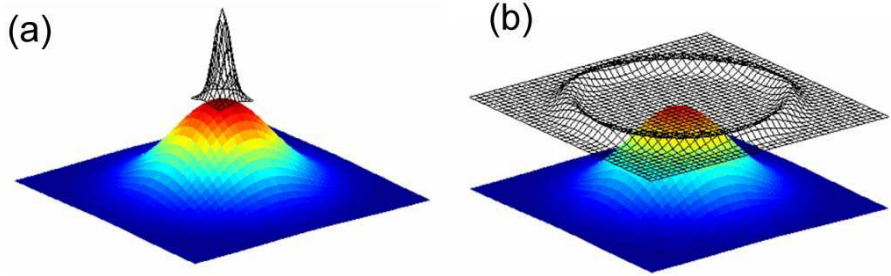}
\caption{Artistic rendition of the operation of the atomic axicon.  A condensate (wavefunction shown in black and white) is initially localized near the origin.  The external potential is shown in false color.  (a) At time $t=0$, the linear potential is turned off, and a blue-detuned optical plug is turned on.  (b)  For times $t>0$ the atoms rapidly ``roll down'' the potential hill, forming a ring-shaped density distribution that expands radially outwards in the $x-y$ plane.
}
\label{Figure 1}
\end{figure}

We created a ring-shaped matter wave output by simultaneously shutting off the quadrupole field and instantly turning on a second 532 nm laser beam focused to the atoms' location that rapidly expelled them.  Hereafter we will refer to this expelling beam simply as "the plug beam".  It propagated along the $z$-axis and had a 70 $\mu$m $1/e^2$ radius (roughly 4 times larger than the cloud radius) and variable power. The entire condensate was instantaneously located on the top of an azimuthally symmetric potential hill with a steep gradient and began to accelerate radially outwards, as illustrated schematically in Figure 1. This strong radial acceleration caused the atom cloud to form an extremely thin, nearly perfectly shaped ring that propagated outward (see Figure 2a for the ring structure and associated movie).

To observe the expanding ring, we waited for a variable time delay   after turning on the plug beam ($\tau$  is hereafter denoted time-of-flight, or TOF).  We then took an absorption image along the $z$-direction using a 200 microsecond probe pulse resonant with the transition from the $F=1$ ground level to the $F'=2$ excited level.   After a few milliseconds, the atoms had been expelled from the region of the laser focus.  After this point, neither the external potential nor the interactions played a significant role (the expansion had lowered the density to the point where interaction effects were negligible).  The subsequent dynamics of the cloud was thereafter purely ballistic.

We observed a narrow ring-shaped atom density distribution (see the movie in Figure 2a).  Its width could only be measured accurately after carefully focusing the imaging system on the falling atom cloud (gravity is parallel to the $z$-axis).  Since atoms gain exactly the same kinetic energy independent of which direction they were expelled, the nearly perfect circular shape was completely independent of minor misalignments of the plug laser or deviations from cylindrical symmetry of the laser focus.

\begin{figure}
\centering
\begin{tabular}{cc}
\begin{minipage}{5in}
\includegraphics[width=0.9\textwidth]{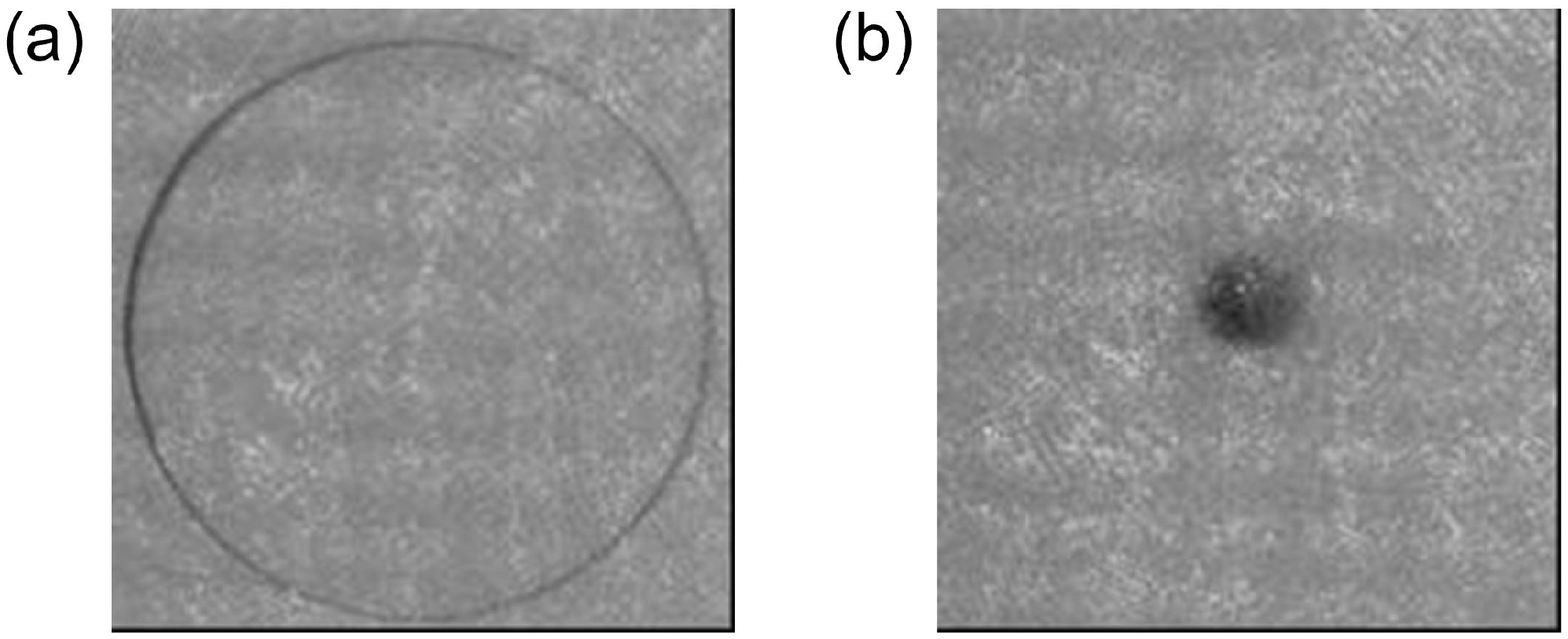} 
\vspace{0.5cm}
\end{minipage}\\
\begin{minipage}{5in}
\includegraphics[width=0.9\textwidth]{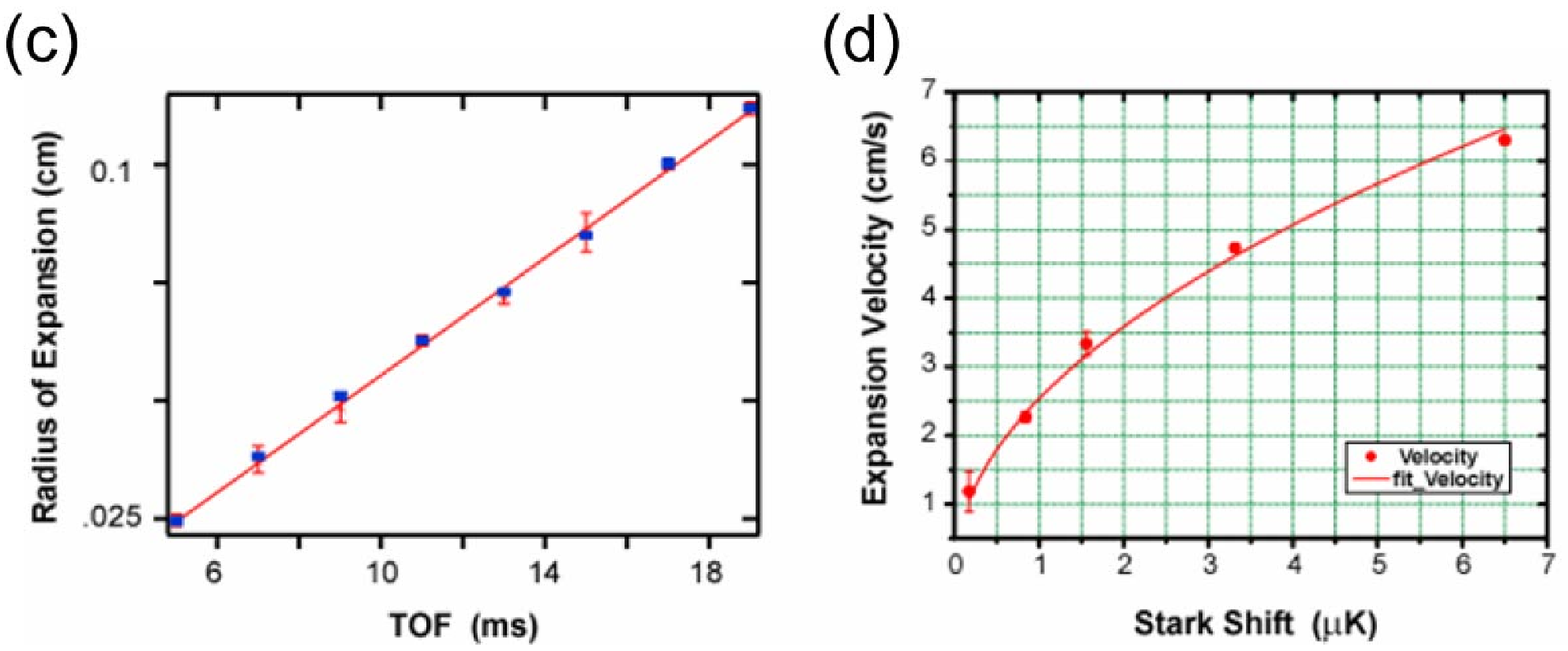} 
\end{minipage}\\
\end{tabular}
\caption{Far field pattern of a conical atom beam.  (a) Movie of the ring expansion and (b)  movie of ordinary BEC expansion [Multimedia available for download at \url{http://www.physics.gatech.edu/chandra/index.htm}]. The frame size is 2.6 mm$\times$ 2.6 mm.  (c) Ring radius versus TOF and (d) velocity of expansion versus the peak AC Stark Shift, proportional to the laser intensity.  The solid curve is a fit to a square root function arising from the energy conservation relation Eqn.\ \ref{eq:vfinal} in the text.}
\label{Figure 2}
\end{figure}

By contrast, the movie in Figure 2b shows the mean-field driven expansion of the BEC in the absence of the plug at a similar time-of-flight.  The mean-field pressure creates a broad distribution of radial velocities whose spread is of the order of the sound velocity in the condensate, about 1 cm/s.  In the presence of the plug, however, this distribution becomes squeezed into a narrow packet traveling at a large mean velocity of up to 6 cm/s.  We observed such thin rings only for Bose-Einstein condensates, where the initial spatial and momentum distributions were simultaneously narrow.  At our longest times-of-flight, these matter wavepackets had a spatial width which was 70 times smaller than the ring radius.  Moreover, apart from an increase in the radius of the ring with time, the shape of the density distribution did not change substantially over the timescale of our observations.

The ballistic nature of the expansion can be clearly seen in Figure 2c for times-of-flight greater than 5 ms.  The maximum of the absorption occurs on a ring whose radius increases linearly with time.  The slope of this line is simply the atomic velocity $\rm{v_f}$ , which we have plotted in Figure 2d as a function of the laser intensity, proportional to the peak AC Stark shift $U$, which was determined from the measured laser power and beam width.  The velocity fits well to a square root function $\rm{v_f}  \propto U^{1/2}$ with the expected prefactor for energy conservation.

We can understand our observations by considering the focused laser beam as a {\em refractive element} for atom optics and by neglecting atom-atom interactions, at least initially. The inhomogeneous dipole force shapes the transverse velocity distribution of the atoms in the following manner.  For a laser beam with $1/e^2$ intensity radius $W$ the AC Stark shift potential is given by $U(\rho) = U_0 e^{-2 \rho^2/W^2}$.  An initially stationary atom which is located at transverse position $\rho$ satisfying $0<\rho<R_\rho$, where $R_\rho$ is the Thomas-Fermi radius of the condensate in the $x-y$ plane, will be accelerated to an asymptotic radial velocity given by the energy conservation relation:
\begin{equation}
\rm{v_f} = \sqrt{\frac{2 U_0}{M}}e^{-\rho^2/W^2} = \sqrt{\frac{2 U_0}{M}}\left[1-\frac{\rho^2}{W^2}+O\left(\frac{\rho^4}{W^4}\right)\right]
\label{eq:vfinal}
\end{equation}
Thus if $R_\rho \ll W$, to lowest order each atom will be given the same positive radial velocity $\rm{v_{f0}} = \sqrt{\frac{2 U_0}{M}}$ independent of its initial location.  The correction terms will shape the final distribution, as we will see later.  In an analogy with conventional optics, we identify the atoms' transverse velocity with the angle of propagation of light rays with respect to an optical axis.  The coordinate along that axis is the equivalent of time in our experiment.  In this picture, our atoms experience the equivalent of a negative axicon lens, which deflects incident light rays through a fixed angle independent of their transverse spatial position measured with respect to that axis.\footnote{It should be noted that there is one difference--the transverse velocity change is only constant for an initially stationary Bose-Einstein condensate, whereas an optical axicon is insensitive to the incident {\em angle} of the light rays as well as their position, with the change in angle being a constant in the paraxial approximation.  For our atoms, this would correspond to a non-zero radial momentum before the plug is applied.}

There have been relatively few realizations of refractive elements for atom optics.  Most atom optics experiments have dealt primarily with {\em thin} lenses, where the change in trajectory during the interaction with the lens element can be neglected.  Kaenders et al.\ have realized a positive axicon for a slow atomic beam by exploiting the linear magnetic potential of a quadrupole magnet, which resulted in a uniform transverse force across the sample that caused the beam to focus on a line \cite{kaen96}.  Since the laser is not pulsed in our experiment, but simply turned on instantaneously, the atoms' velocity distribution is significantly modified {\em during} the interaction with the beam.  Thus we have realized a {\em thick} atom lens, which does not require the force to be constant in space as it was for the thin lens implemented by Kaenders et al.  The conical feature of the final velocity distribution arises simply from energy conservation and an extremely narrow spatial distribution of atoms $R_\rho \ll W$, and thus does not depend on the details of the forces which are applied.  The implementation of thick lenses is greatly facilitated using Bose-Einstein condensates since the atomic dynamics are intrinsically slow.  It is therefore straightforward to engineer a long interaction time with the lens element, which opens up new possibilities for realizing refractive elements for atom optics \cite{bloc01optics,arno04}.

The radius of the ring is determined by the peak AC Stark shift, which determines the energy gained by atoms in traversing the plug potential.  This is the equivalent of the cone angle of the axicon lens, which determines the deflection angle.  However, the radial distribution of the atom cloud within the ring itself is more complex, and one must in general consider both the effects of interactions as well as the potential due to the lens.  We have measured this distribution for various times of flight and the data are shown in Figure 3 for our maximum plug Stark shift of $k_B \times $ 6.5 $\mu$K.  Before application of the plug, the Thomas-Fermi density profile of the atoms is given by $n_i (\rho ,z) = n_0 (1 - \sqrt {R_\rho  ^2  - \rho ^2  - 4z^2 } )$, where $n_0$ is the peak density.  After the interaction, the final density profile is given in cylindrical coordinates by $n_f (\rho ,\phi ,z)$, where $\phi$ is the azimuthal angle.  We measure the column density $\tilde n_f (\rho ,\phi ) = \int {n_f (\rho ,\phi ,z)dz}$.  Due to the cylindrical symmetry of the problem, we can define an azimuthally averaged density profile $\tilde n_f (\rho ) = 2\pi \rho \left( {{1 \over {2\pi }}\int {\tilde n_f (\rho ,\phi )d\phi } } \right)$, from which the normalization condition $\int\limits_0^\infty  {\tilde n_f (\rho )d\rho } = N$ results, where $N$ is the number of atoms.

In Figure 3a we have plotted the measured atom radial density  for a succession of times-of-flight from 5 ms to 19 ms.  The average allowed us to account for slight misalignments between the plug and the Thomas-Fermi distribution of the atoms, which caused some azimuthal variations in the atom density.  Each 1-dimensional data set was obtained by averaging the two-dimensional image using the following procedure.  A 1 pixel-thin slice was taken through the image after it was rotated through an angle $\theta$ about the ring center, which could be readily determined to within the pixel size, about 5 $\mu$m.  For each slice, we folded the data about the center to obtain two data sets corresponding to angles $\theta$ and $\theta+\pi$, both starting at $\rho = 0$.  The process was repeated 180 times with 1 degree rotation increments and the final radial distribution for each image was the sum of all the data sets.  For all times-of-flight, the distribution clearly has an asymmetric shape, with a long tail extending toward smaller radii and a relatively sharp cutoff at larger radii.  We quantified the asymmetry by defining the half-width at half-maximum (HWHM) for the distribution with respect to its peak, defined as $\rho =\rho_0$.  This allowed us to define a backward and forward HWHM, $W_{back}$ and $W_f$, respectively, corresponding to   $\rho < \rho_0$ and $\rho  >  \rho_0$, respectively.  These are plotted in Figure 3b, and clearly show that $W_{back} > W_f$ for the majority of the data points.

\begin{figure}[htb]
\centering\includegraphics[width= 0.75\textwidth]{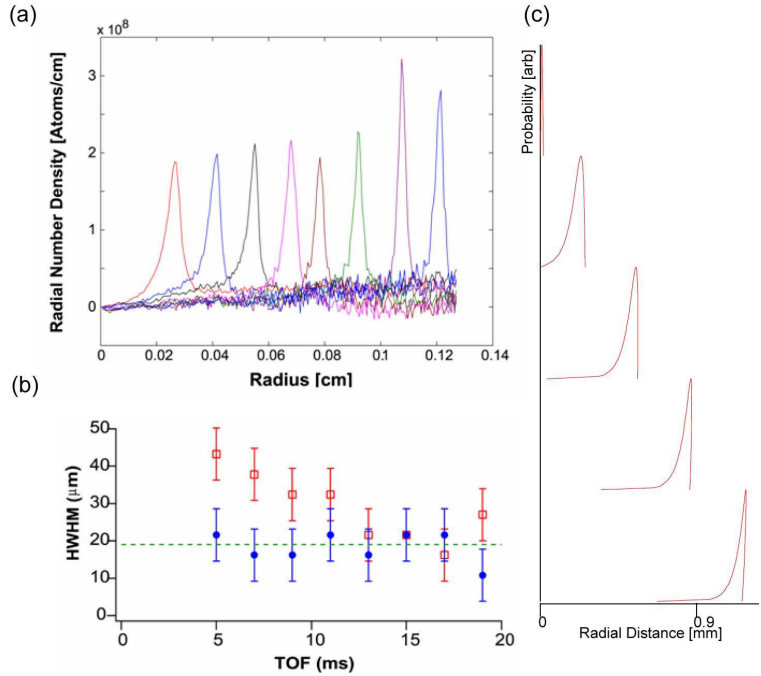}
\caption{Spatial distribution of the ring.  Data were taken at the maximum plug Stark shift of $k_B \times $ 6.5 $\mu$K.(a)  Radial number density  of atoms for TOF from 5 to 19 ms in increments of 2 ms, showing the outward radial evolution.  (b)  Spatial Half-Width at half-maximum (HWHM) for forward width $W_f$ (solid circle) and backward width $W_{back}$ (square). (c) Time evolution of a classical ``wavepacket'' simulation of our experiment, as described in the text.  The probability density is shown for times 0, 5, 10, 15 and 20 ms, and confirms the temporal focusing observed in (b).}
\label{Figure 3}
\end{figure}

We can explain our results in terms of a simple atom-optical picture in which the plug laser acts as a lens that focuses the atomic distribution.  To understand the physics, we concentrate on two groups of atoms--group A initially located at $\rho $ close to 0 and group B located at $\rho $ close to $R_\rho$, the Thomas-Fermi radius of the BEC.  At $t=0$, atoms in group A experience a negligible force since the gradient in the laser intensity is close to zero.  Therefore, they remain momentarily stationary.  However, atoms in group B experience an immediate force due to the finite gradient, and begin moving radially outward.  This leads to a momentary spreading of the wavepacket.  This spreading is aided by atom-atom interactions, which act to reduce the atom density $n$ further through the positive mean-field pressure $\propto n^2$ \cite{pita03book}.  However, for longer times, group A atoms begin to move outward as well.  They acquire more kinetic energy than group B atoms due to their larger initial potential energy, as expressed in Eqn.\ \ref{eq:vfinal}.  Eventually, group A will catch up with group B at some intermediate time $\tau_f$, which we denote the {\em focusing time}, when the spreading is reversed.  This focusing time is, typically, considerably larger than the average transit time of atoms $\tau_{trans}$ through the beam.  The key is that for times $\tau_{trans} < t \leq \tau_f$ the atoms' velocity is close to the asymptotic value given by Eqn.\ \ref{eq:vfinal}, but their {\em spatial} distribution has not yet equilibrated.  Moreover, by this time the axicon has reduced the atom density by spreading the atoms onto a ring, and interactions have no further effect on the propagation, which is governed purely by linear optics.  Finally, at very long times $t \gg \tau_f$, the wavepacket should defocus.  That is, it will resemble a map of the asymptotic velocity distribution, with fast atoms in front, and slow atoms behind.

Our experimental observations lie precisely in this intermediate timescale $t \sim \tau_f$.  Thus we observe no increase in the forward width $W_f$ and a reduction in the backward width $W_{back}$.  The latter provides evidence of the temporal focusing of the wavepacket.  Further evidence was provided by classical simulations which we have performed based on the specific parameters of our experiment.  We numerically computed $n_{max}$ single atom trajectories in the presence of the plug for times $0<t<t_{sim}$.  We chose various starting locations $0< \rho_{i,k}  < R_\rho$ at time $t=0$, where $k=1,2,...n_{max}$, and computed a sequence of final locations $\rho_{f,k}$ at $t=t_{sim}$.  The resulting classical ``wavepacket'' was constructed by assigning to each $\rho_{f,k}$ a probability density weighted in proportion to the initial number of atoms located at $\rho_{i,k}$, i.e.,$\propto 2 \pi \rho_{i,k} \int {n_i (\rho_{i,k} ,z)dz}$, the column density in the trap.  The integration limits used are $z = \pm \frac{1}{2} \sqrt{R_\rho^2-\rho^2}$, located on the Thomas-Fermi surface of the cloud.  The result is shown in Figure \ref{Figure 3}c for $n_{max} = 200$ trajectories and times $t_{sim} = 0,5,10,15,20$ ms.  The simulated wavepacket shows a strong resemblance to the data in Figure \ref{Figure 3}a.  In particular, we observe an asymmetric distribution with a shrinking tail, i.e., $W_{back}$ decreases with time.  A close examination of the final classical wavepacket shows that it is in fact multi-valued at the leading edge, an indication that many classical trajectories have crossed one another, an effect confirmed by a detailed examination of the individual trajectories used to construct the wavepacket.  At this point the classical analysis begins to break down, and one requires a full quantum treatment.  Nonetheless, this simple simulation confirms the temporal focusing of the axicon with a time on the order of, or close to 20 ms.

This ``ray optics'' picture cannot accurately be used to compute the atom distribution near the temporal focusing point.  However, it is intuitively clear why the forward width $W_{f}$ in our measurements remains stationary--atoms simply pile up from behind, and do not spread out until a much longer time $t \gg \tau_f$, when the different momentum components separate again.  Unfortunately, we could not observe the ring at larger TOF as it exceeded the camera sensor size\footnote{Demagnification of the image would have allowed us to observe the ring at larger TOF, but at the cost of reduced spatial resolution, which would compromise the width measurement.}.  Although our experiments do not measure the asymptotic momentum distribution, we can make a rough estimate of the M$^2$, or quality factor for our atom beam, which measures the phase space volume occupied.  From the reduction of $W_{back}$ with time in Figure \ref{Figure 3} we estimate a velocity width of order $\delta v_r \sim 2$ mm/s.  The measured spatial width near the temporal focus is no larger than 20 $\mu$m, and therefore, M$^2 = \frac{2 M}{\hbar} \delta r \times \delta v_r \leq 30$.  While not yet at the diffraction limit, we note that typical values for RF outcoupled atom lasers are in the range of $\sim 10$, not far from our result \cite{riou06}.

In a further analogy with optics, our experiment may be described as a ``cavity-dumped'' atom laser, since the entire condensate is emitted as a single matter wave pulse.  There are close analogies between our work and a host of experiments that extract atoms {\em slowly} from a BEC, where the inhomogeneous mean field interaction between outcoupled atoms and the stationary condensate affects the spatial profile of the laser.  For example, RF coupling results in a roughly constant energy but a broad transverse distribution due to the mean-field \cite{mewe97,leco01}, which can result in distortion of the spatial mode due to gravitational sag \cite{riou06}.  By contrast, our conical matter wave is extremely symmetric, and is relatively free of mode distortions due to the fact that gravity plays no important role in the extraction process.  In this regard, our method is a significant step forward in comparison with RF techniques, especially when outcoupling away from the Thomas-Fermi surface.  The conical mode shape could be useful for a number of atom-optical applications where the unique spatial mode can be exploited, for example, to form Bessel atomic beams that do not spread out in time.  We plan to explore this avenue in the future.

\begin{figure}[htb]
\centering\includegraphics[width= 0.75\textwidth]{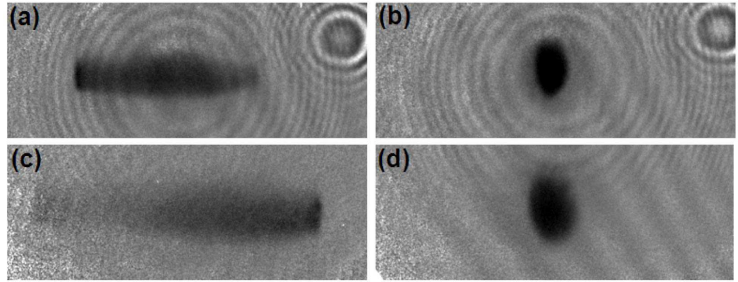}
\caption{Side view of expanding ring shown at 10 ms (top: a,b) and 15 (below: c,d) ms for both ring shaped expansions (left: a,c) as well as ordinary BEC expansion (right: b,d).  Each image is 0.8 mm $\times$ 2.2 mm.}
\label{Figure 4}
\end{figure}

While the plug created a narrow radial wavepacket, we found that the atoms' velocity along the $z$-direction was almost completely unaffected.  Due to the long confocal parameter of 2.8 mm, there is a negligible variation of the laser intensity and therefore no force exerted along that direction over the relevant experimental timescales.  In Figure 4 we have observed the atoms after 10 and 15 ms TOF imaged along one of the radial directions.  The ring of atoms appears in this view as a band with a considerable axial width, comparable to that of an ordinary BEC expansion in the absence of the plug, as shown in Figure 4(b,d).

\section{Theory}

While much insight was gained by a purely classical simulation, the full dynamics requires a quantum mechanical treatment.  To this end, we have simulated the dynamics of the BEC in the presence of the plug in cylindrical coordinates $\rho$  and $z$, taking advantage of the azimuthal symmetry of the problem.  We describe the condensate by the order parameter $\psi \left( {\rho ,z,t} \right)$, normalized such that $n\left( {\rho ,z,t} \right) = N\left| {\psi \left( {\rho ,z,t} \right)} \right|^2 $. The condensate dynamics is governed by the Gross-Pitaevskii equation

\begin{equation}
i\hbar \frac{\partial }
{{\partial t}}\psi \left( {\rho ,z,t} \right) = \left( { - \frac{{\hbar ^2 }}
{{2m}}\nabla ^2  + V\left( {\rho ,z,t} \right) + Ng\left| {\psi \left( {\rho ,z,t} \right)} \right|^2 } \right)\psi \left( {\rho ,z,t} \right)
\label{eq:gpe}
\end{equation}
where $N$ is the number of atoms, $g = 4\pi \hbar ^2 a/m$ is the mean field interaction strength, $m$ is the atomic mass, $a$ is the scattering length, and $V\left( {\rho ,z,t} \right)$ is the combined external potential applied by the magnetic trap and the plug beam.  We assume the magnetic trap is turned off and the plug beam is turned on instantaneously at time $t=0$.  Explicitly, we write the potential as

\begin{equation}
V({\mathbf{r}},t) = \left\{
\begin{array}{ll}
\mu_M B_{\rho}' \sqrt {\rho ^2  + 4z^2 } & ,t < 0 \\
U_0  \exp{ \left( - \frac{2\rho ^2}{W^2} \right)} & ,t > 0 \\
\end{array}
\right.
\end{equation}
where $\mu _M  \approx 2\pi \hbar  \times 0.7$ MHz/G is the atomic magnetic moment, $B_{\rho} '$ is the radial magnetic field gradient, $U_0$ is the Stark-shift at the center of the plug, and $W$ is the $1/e^2$ radius of the plug.  At $t=0$, we assume the condensate is in the ground state of the magnetic trap with the plug turned off.  The ground state is itself calculated by evolving Eqn.\ \ref{eq:gpe} in imaginary time.  The dynamics induced by the plug is then found for $t>0$ by integrating Eqn.\ \ref{eq:gpe} in real time.  The grid size, spatial resolution, and time step were varied to ensure convergence.  To ease the computational burden, the simulation was done for a smaller atom number than the experiment but with the same magnetic field gradient (see Table 1).  We used approximately the same values of the ratio of plug potential to chemical potential ($\mu _a$), and the ratio of plug radius to Thomas-Fermi radius of the condensate as in the experiment.

\begin{table}[htb]
\centering\caption{Comparison of theoretical and experimental parameters for ring shaped expansion.}
\begin{tabular}{lp{0.5in}|lp{0.5in}|lp{0.5in}|lp{0.7in}|lp{1in}|lp{0.3in}}
\hline
 & $W$ ($\mu$m)   & $R_{\rho}$ ($\mu$m)  & $V_0 /k_B$ ($\mu$m)  & $\mu_a/k_B$ ($\mu K$) & $N$   & $B_{\rho}'$ (G/cm)    \\ \hline
Theory  & 23.3  & 6.4   & 2.2 $\mu$ K   & 0.255 $\mu$ K & $4.9\times10^4$   &  12 \\
Experiment & 70 & 18    & 6.4 $\mu$ K   & 0.7 $\mu$ K   & $3\times10^6$     &  11 \\
\hline
\end{tabular}
\end{table}

In Figure 5 we show the numerical results of the expansion, which provide key qualitative insights into our data.  The initial density profile is shown in Figure 5a, and closely follows the Thomas-Fermi distribution.  After applying the plug potential for a time 1.3 ms, the simulation was stopped and the final density distribution computed as shown in Figure 5b.  Figure 5c shows the corresponding radial momentum distribution of the wavefunction.  By this time the density had been lowered by a factor of 200 due to the expansion caused by mean-field repulsion of atoms as well as the radial force of the plug potential.  Moreover, at this point in time, the peak of the atom density has moved beyond the $1/e^2$ radius of the plug potential.  Therefore, the subsequent expansion of the cloud should be mostly ballistic, with only small corrections to the momentum distribution expected.

The density profile in Figure 5b shows that the cloud has expanded considerably in the $z$-direction.  Moreover, in the radial direction, the atoms are being forced outward and form a steep front at $\rho \cong 32$ $\mu$m.  However, a spatial tail of the distribution still exists for smaller $\rho$.  Our data in Figure 3a for 5 ms and at short TOF bears a striking similarity to the spatial distribution in Figure 5b.  Figure 5c shows the velocity distribution of the atoms and confirms that the majority of the atoms occupy a very narrow peak in momentum space near 38 mm/s, corresponding to a kinetic energy of 1.95 $\mu$K, about 90\% of the peak plug potential.  This is consistent with our observations in Figure 3b of an extremely narrow velocity spread for the expanding ring.  The velocity width (HWHM) computed from the simulation is about 600 $\mu$m/s, about a factor of 3 larger than the Heisenberg limited width of 215 $\mu$m/s for a cloud of radius $R_{\rho} = 6.4 \mu$m.  For comparison we have also shown the momentum distribution of a mean-field driven expansion in the absence of the plug.  To understand the role that mean-field effects play on the ring formation, we performed similar calculations for an expansion time of 1.3 ms using the same initial wavefunction but with the mean-field term in the Gross-Pitaevskii equation set to zero.  This represents the purely ``linear'' evolution of the atomic wavepacket.  The resulting momentum distribution is also shown in Figure 5c, and is qualitatively very similar to the nonlinear evolution.  A similar sharp peak with a narrow velocity width is formed, demonstrating that the dynamics are largely insensitive to the effects of interactions.

Two other features emerge from the simulations--in both the interacting and non-interacting cases in Figure 5c there is a long tail extending to low velocities, most likely due to the slow evolution of atoms initially located very close to $\rho = 0$.  These atoms will catch up with the peak of the wavepacket only at later times.  Moreover, it is intriguing to note the presence of smaller, ``satellite'' momentum peaks in both simulations which appear to be interference phenomena.  To explore these further, we extended the non-interacting calculation to the longer time of 1.8 ms.  This calculation proved easier than the fully interacting case due to the larger grid size used.  Due to the similarity between the curves in Figure 5c, the linear evolution provides a useful guide to the dynamics.  An expanded view of the momentum peaks is shown in Figure 5d at both 1.3 and 1.8 ms for the linear evolution.  The interference peaks appear to increase in amplitude with time, and coincide with a reduction of the low velocity tail.  It is clear that the radial phase profile is not flat, and there may be differences between the interacting and non-interacting cases at longer times.  Future work will explore these interferences in detail, with an eye toward understanding their effect on atom-optical applications of the axicon lens, where the phase profile is of importance.

\begin{figure}[htb]
\centering\includegraphics[width= 0.8\textwidth]{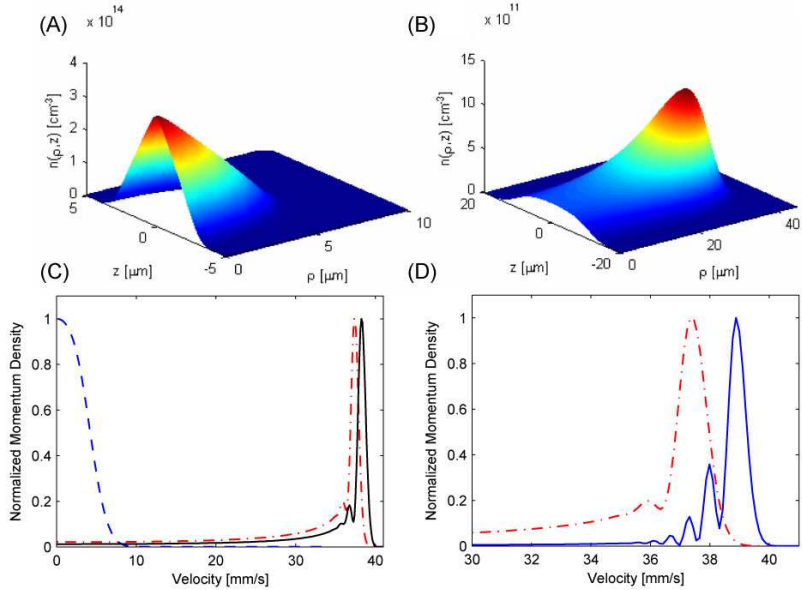}
\caption{Numerical simulation of ring shaped expansion.  Computed density profiles in the $\rho$-$z$ plane for (a) the initial Thomas-Fermi distribution and (b) at 1.3 ms, after the atoms have mostly traversed the region containing the plug.  In (c) we show the velocity distribution of the wavefunction of (b) (solid black) as well as that obtained for a purely linear temporal evolution, i.e., by turning off the nonlinear term in the Gross-Pitaevskii equation during the expansion (dot-dashed red).  For comparison, we also show the velocity distribution obtained by mean-field expansion without the plug (dashed blue).  In (d) we examine the region near the velocity peak for the purely linear expansion at times 1.3 (dot-dashed red) and 1.8 ms (solid blue).  For each curve, the velocity distribution is normalized to its peak value.}
\label{Figure 5}
\end{figure}

\section{Conclusion}

In summary, we have demonstrated a conical lens for atoms that produces an extremely narrow velocity distribution, which could have applications to interaction-free matter wave manipulation.  To our knowledge, this is the first observation of the far field pattern of such a conical lens in atom optics.  Our results point to the power of using Bose-Einstein condensates, where one can readily apply thick refractive elements to shape the spatial and momentum distributions.

\section{Acknowledgment}
This work was supported by the DoE, ARO, NASA and Georgia Tech.

\end{document}